\documentclass{aa}
\usepackage{graphicx}
\usepackage{comment}
\usepackage{multirow}
\usepackage{upgreek}
\usepackage{hyperref}
\usepackage{pifont}
\usepackage{pdfpages, caption}
\DeclareUnicodeCharacter{00A0}{ }
\usepackage{txfonts}
\begin{document} 

   \title{A re-analysis of equilibrium chemistry in five hot Jupiters}

   \author{E. Panek \inst{1} \fnmsep\thanks{emilie.panek@iap.fr}
        \and
          J-P. Beaulieu \inst{1,2}
          \and
          P. Drossart \inst{1,7}
          \and
          O. Venot \inst{3}
          \and
          Q. Changeat \inst{4,5}
          \and
          A. Al-Refaie \inst{5}
          \and
          A. Gressier \inst{4,1,6,7} 
          }

   \institute{Institut d'Astrophysique de Paris (CNRS, Sorbonne Université), 98bis Bd Arago, 75014 Paris, France
         \and
             School of Physical Sciences, University of Tasmania, Private Bag 37 Hobart, Tasmania 7001, Australia
        \and
            Universit\'e Paris Cit\'e and Univ Paris Est Creteil, CNRS, LISA, F-75013 Paris, France
        \and
            Space Telescope Science Institute (STScI), 3700 San Martin Dr, Baltimore MD 21218, USA
        \and 
            Department of Physics and Astronomy, University College London, London, UK
        \and
            LATMOS, CNRS, Sorbonne Université/UVSQ, 11 boulevard d’Alembert, F-78280 Guyancourt, France
        \and
            LESIA, Observatoire de Paris, Université PSL, CNRS, Sorbonne Université, Université de Paris, 5 place Jules Janssen, 92195 Meudon, France\\
            }

   \date{Received 2023; accepted 2023\\}


 
  \abstract
   {}
   {Studies of chemistry and chemical composition are fundamental to exploring the formation histories of planets and planetary systems. We propose having another look at five targets to better determine their composition and the chemical mechanisms taking place in their atmospheres. We present a re-analysis of five hot Jupiters, combining multiple instruments and using Bayesian retrieval methods. We compare different combinations of molecules present in the simulated atmosphere and various chemistry types, as well as a range of cloud parametrizations. Following up on recent studies questioning the detection of Na and K in the atmosphere of HD 209458b as being potentially contaminated by stellar lines (when present), we study the impact on other retrieval parameters that may lead to misinterpretations of the presence of these alkali species.}
   {We used spatially scanned observations from the grisms G102 and G141 of the Wide Field Camera 3 (WFC3) on the Hubble Space Telescope, with a wavelength coverage of $\sim$0.8 to $\sim$1.7 microns. We analyzed these data with the publicly available Iraclis pipeline. We added data from Space Telescope Imaging Spectrograph (STIS) observations to increase our wavelength coverage from $\sim$0.4 to $\sim$1.7 microns. We then performed a Bayesian retrieval analysis with the open-source TauREx using a nested sampling algorithm. We carried out the retrieval, taking into account molecular abundances that vary freely and then with equilibrium chemistry. We explored the influence of including Na and K on the retrieval of the molecules from the atmosphere.}
   { Our data re-analysis and Bayesian retrieval are consistent with previous studies, but we do find small differences in the retrieved parameters. After all, Na and K have no significant impact on the properties of the planet atmospheres.
   Therefore, we present here our new best-fit models, taking into account molecular abundances that are allowed to vary freely as well as the equilibrium chemistry. This work is a preparation for a future addition of a more sophisticated representation of the chemistry involved, while taking into account disequilibrium effects such as vertical mixing and photochemistry.\\}
   {}


   \maketitle
%

\section{Introduction}
\label{section:1intro}

{\renewcommand{\arraystretch}{1.4}
\begin{table*}[h!]
\caption{List of the physical parameters of the planetary systems studied in this paper. When two publications are indicated, it refers to both the associated stellar parameters and planetary parameters published. }
\label{tab:parameters}
\centering
\begin{tabular}{ccccccc}
\hline\hline
Planets & $T_*$ & $R_*$ & $T_{eq}$ & $R_p$ & $M_p$ & References\\\hline 
HAT-P-12b & 4650 & 0.70 & 975 & 0.949 & 0.211 & \cite{Ozturk}\\
HD 209458b & 6026 & 1.19 & 1448 & 1.39 &0.73 & \cite{Stassun}\\
WASP-6b & 5380 & 0.82 & 1150 &1.03 &0.37 & \cite{Stassun}\\
WASP-17b & 6548 & 1.57 & 1740 & 1.89 & 0.51 & \cite{Stassun2}, \cite{Barstow}\\
WASP-39b & 5326 & 1.01 & 1120 & 1.27 & 0.28 & \cite{Stassun2}, \cite{Barstow}\\
\hline\hline \\
\end{tabular}
\end{table*}
}

The chemical composition of an exoplanetary atmosphere can provide valuable information on the physical conditions and evolutionary history of the planet \citep{Madhusudhan, Mordasini_2016, Brewer_2017,Eistrup, Turrini_2, Turrini_1, Lothringer}. One of the most reliable methods for studying the composition of an exoplanet atmosphere is the method of transmission or transit spectroscopy \citep{Seager, Redfield, Snellen,Burrows}.

The principle of transit spectroscopy is to measure spectral variations of transit depth. Considering opacities of constituent molecular species in the atmosphere are wavelength-dependent, it is an evaluation of atmospheric composition. As the planet passes in front of its host star, a small portion of starlight travels throughout the planet's atmosphere: molecules then absorb the light of a certain wavelength, while light of other wavelengths can pass through the atmosphere without being absorbed.

This transmission spectroscopy method has been used to study all types of planetary atmospheres, from hot Jupiters \citep{Sing_2015,Barstow, Tsiaras_2018, Pinhas, Carter2020, Skaf, Pluriel, Roudier_2021, Saba,Changeat_2022, Edwards_2022} to super-Earths \citep{Kreidberg_2014, Tsiaras_k218, Edwards, Swain, Mugnai, Libby_Roberts}.

The motivation behind this paper is to prepare for the ESA/Ariel mission \citep{Tinetti, Tinetti_redbook} and its statistical sampling of exoplanets from transit spectroscopy. The methods and data analysis presented here will be similar to those of future Ariel spectral retrievals.\\

Low-resolution data allow us to derive a limited amount of information about the atmosphere, which is why models often use simplistic assumptions such as the use of constant chemical profiles throughout the atmosphere. Although it is not representative of the entire atmosphere, the observations currently available account for only a small pressure range, which undergoes very few variations in terms of abundance. However, the arrival of more powerful space telescopes such as JWST and Ariel encourages the development and use of more realistic chemical models of the abundance variations that take place in the atmosphere of an exoplanet. This is why we are interested in the study of different chemical models applied to hot Jupiters, using chemical equilibrium or disequilibrium schemes in the present study.

Here, we study five hot Jupiters that are inflated due to their proximity to their host star, using Hubble Space Telescope (HST) data. For years, HST has served has allowed for great progress to be made in the characterization of the atmospheres of exoplanets. More specifically, we used data from the Wide Field Camera 3 (WFC3) and the Space Telescope Imaging Spectrograph (STIS), which have been used in many studies of planetary atmospheres in recent years \citep{Gibson, Sing_2015, Tsiaras_2018, Gressier}.\\

The choice of the targets is based on the sample of planets presented in \cite{Sing_2015}. We decided not to take the entire sample of ten planets and we preferred to reduce it to five targets. As we are interested in studying the signatures of sodium and potassium in the spectra, we set aside HD 189733b, WASP-12b, and WASP-19b, for which the signatures of these species are not visible. We also wanted to see whether disequilibrium chemical schemes can be used in an inversion method to find more precise and more realistic parameters of atmospheres. The more promising targets featured here are: HAT-P-12b, HD 209458b, WASP-6b, WASP-17b, and WASP-39b. \\

HD 209458b was the first planet on which molecular absorption signatures were detected \citep{Charbonneau}, including the signature of neutral sodium, which was later confirmed with the same data by \cite{Sing_2008}. Ground-based observations have also confirmed the sodium signatures with higher resolution \citep{Snellen,Albrecht,Jensen_2011}. However, more recent studies have raised doubts on the signatures of sodium as well as other species, such as potassium, \citep{Casasayas,Casasayas2,Morello_Na}, attributing the feature to the Rossiter-McLaughlin effect. This is why we discuss the effect of the signatures of sodium and potassium on the parameters of retrievals on the spectrum of HD209458b and on the spectra of the four other targets presented in this study (Section \ref{section:3.4contrib_Na_K}).\\

In Section \ref{section:2data} we present our five targets, along with the data reduction process and the different steps that follow the Iraclis software. Then, we compare our reduced data sets with other reductions in the literature. In Section \ref{section:3method} we explain our methodology: the TauREx program and our TauREx set-up. We also describe how we established our different chemistry types and cloud parametrizations. We also talk about the effect of the sodium and potassium contributions. In Section \ref{section:4results} we review our results planet-by-planet. Then, in Section \ref{section:5discussion}, we discuss the interpretation of the results, the limitations of our current models, and the possibilities for future works.
Finally, we draw our conclusions in Section \ref{section:6:conclusion}.


\section{Data}
\label{section:2data}

\subsection{Targets}

We selected a sample of five planets : HAT-P-12b, HD 209458b, WASP-6b, WASP-17b, and WASP-39b. All of them are presented in \cite{Sing_2015}. They are hot Jupiters, so they have strong signal-to-noise ratios (S/Ns), but they nevertheless present a variety of different parameters: radii ranging from a 0.9 to 1.9 Jupiter radius and atmospheric temperatures varying from 1000K to 1700K. The physical characteristics of these planets used in this study can be found in Table \ref{tab:parameters}. These targets have also been the subject of several previous publications with which we have compared our work \citep{Sing_2015, Wakeford_2017, Carter2020, Saba, Wong}, as explained in Section \ref{section:2-3comp}.\\

We performed a data reduction analysis using the Iraclis software \citep{Tsiaras_2016a, Tsiaras_2016b, Tsiaras_2018}, then we executed a Bayesian retrieval analysis with the TauREx 3.1 program \citep{AlRefaie,Al_Refaie_chemcomp}. The retrieval technique has proven its effectiveness in inverting spectra and deducing the parameters of the atmospheres \citep{Lee, Waldmann,Irwin_nemesis,Molliere_2019_pRT}.
The improvements of the data reduction analysis performed in this study enables a better accuracy in chemical composition retrieved, which lead to discussion on the potential chemical equilibrium composition in the atmosphere of these objects.\\

Here, we use HST STIS and WFC3 data on each planet.
There are Spitzer or ground observations available for some of our targets \citep{Beaulieu, Saba}, but we preferred not to include them to maintain greater consistency across the analysis.\\

For each planet, we used several transit observations with the HST STIS grisms 430L and 750L that ranges from 0.29$\upmu$m to 1.027$\upmu$m, as well as the HST WFC3 grisms G102 and G141 that ranges from 0.8$\upmu$m to 1.7$\upmu$m. All information concerning the observations is given in Table \ref{tab:params_observations}.

{\renewcommand{\arraystretch}{1.4}
\begin{table*}
\caption{Details on observations used in the study.}
\label{tab:params_observations}
\centering
\begin{tabular}{|c|c|c|c|c|c|}
\hline\hline
Planets & Instrument & Grism & Date & Proposal ID & Proposal PI\\\hline 
\multirow{5}{*}{HAT-P-12b}& \multirow{3}{*}{HST STIS} & G430L & 2012/05/26 &\multirow{3}{*}{12473} & \multirow{3}{*}{Sing}\\
& & G750L & 2012/05/30 & & \\
& & G430L & 2012/09/19 & & \\
\cline{2-6}
& \multirow{2}{*}{HST WFC3} & G141 & 12/12/2015 & \multirow{2}{*}{14260} & \multirow{2}{*}{Deming} \\
& & G141 & 31/08/2016 &  & \\ \hline
\multirow{9}{*}{HD 209458b} & \multirow{8}{*}{HST STIS} & G430L & 2003/05/03 &\multirow{4}{*}{9447} &\multirow{4}{*}{Charbonneau}\\
& & G750L & 2003/05/31& &\\
& & G430L & 2003/06/25 & &\\
& & G750L & 2003/07/05& &\\
\cline{5-6}
& & G750M & 2000/04/25&\multirow{4}{*}{8789} & \multirow{4}{*}{Brown}\\
& & G750M & 2000/04/28 & &\\
& & G750M & 2000/05/05 & &\\
& & G750M & 2000/05/12 & &\\ \cline{2-6}
 & HST WFC3 &G141 & 2012/09/25 &12181 & Deming\\ \hline
 \multirow{4}{*}{WASP-6b} & \multirow{3}{*}{HST STIS} & G430L & 2012/06/10 & \multirow{3}{*}{12473}& \multirow{3}{*}{Sing}\\
 & & G430L & 2012/06/16 & & \\
 & & G750L & 2012/07/23 & & \\\cline{2-6}
 & HST WFC3 & G141 & 2017/05/06 & 14767 & Sing \\ \hline
\multirow{6}{*}{WASP-17b} & \multirow{3}{*}{HST STIS} & G430L & 2012/06/08 & \multirow{3}{*}{12473}& \multirow{3}{*}{Sing} \\
& & G430L& 2013/03/15 & &\\
& & G750L& 2013/03/19 & & \\\cline{2-6}
& \multirow{3}{*}{HST WFC3} & G102 & 2017/06/16 & \multirow{3}{*}{14918}& \multirow{3}{*}{Wakeford} \\
& & G102 & 2017/09/25 &  &  \\
& & G141 & 2012/07/23 & & \\\hline
\multirow{6}{*}{WASP-39b} & \multirow{3}{*}{HST STIS} & G430L& 2013/02/08 & \multirow{3}{*}{12473}& \multirow{3}{*}{Sing} \\
& & G430L& 2013/02/12 & & \\
& & G750L& 2013/03/17 & & \\\cline{2-6}
& \multirow{3}{*}{HST WFC3} & G102 & 2016/07/07 & 14169 & Wakeford \\
& & G141 & 2016/08/29 & 14260 & Deming \\
& & G141 & 2017/02/07 & 14260 & Deming \\
\hline\hline
\end{tabular}
\end{table*}
}

\subsection{HST WFC3 data reduction}

We performed a re-analysis of the HST WFC3 data with the Iraclis tool (see \citealt{Tsiaras_2016b, Tsiaras_2016a, Tsiaras_2018}). Raw data are available on the Mikulski Archive for Space Telescopes (MAST). Iraclis is an open source pipeline\footnotemark[1] for data reduction and calibration of HST WFC3 data sets. It follows multiple steps, as follows: 
reduction and calibration of each image;
extraction of the 1D spectra from each image;
white light curves fitting;
spectral light curves fitting.\\

\footnotetext[1]{https://github.com/ucl-exoplanets/Iraclis}

We consider limb-darkening coefficients as constant because of degeneracies with other transit-shape parameters. These degeneracies are especially important in our case because of the gaps that periodically occurs in HST data.
Before using Iraclis, we calculated these limb darkening coefficients for each star and in the wavelength band that interests us. We used ExoTETHyS \citep{Morello}, especially the Stellar Atmosphere Intensity Limb (SAIL) subpackage to account for this limb darkening effect. We choose a stellar model based on the ATLAS grid, and we adopted the Claret 4 coefficients limb-darkening law described in \cite{Claret_2000} .\\

After calculating the limb-darkening coefficients, we used Iraclis \citep{Tsiaras_2018, Tsiaras_2016b, Tsiaras_2016a}. This is a pipeline developed to reduce the HST data of the WFC3 instrument for transiting exoplanets, when the scanning mode of the telescope is used.
This method of reduction has been explained in detail in \cite{Tsiaras_2018}, \cite{Saba}, \cite{Mugnai}, \cite{Guilluy}.
We propose here to make a short overview of the different steps of this method.
First, we use Iraclis to apply the standard steps of data reduction: bias correction, zero-read correction, dark current subtraction, detector gain variation correction, flat field correction, and bad pixel correction, as well as cosmic rays corrections.\\

Once these reduction and calibration steps have been performed, the next step is the extraction of the 1D spectra from each image, which gives the wavelength-dependent light curves. We extracted the white light curve, a broad band of wavelengths that includes most of the observed starlight, and the spectral light curves, with the light decomposed on several narrower wavelength bands. \\

Afterwards we proceed to fit the light curves with a transit-shape model using a Markov chain Monte Carlo (MCMC) method implemented in emcee \citep{Foreman-Mackey}.The transit model used is well described in \cite{Tsiaras_2016a}. The parameters fitted here are: $n^{for}_w$ is the normalisation factor which accounts for the scanning direction; $r_{a1}$ is the slope of the linear long-term ramp; $for_{rb1}$ and $for_{rb1}$ are the coefficients of the exponential short-term ramp for the first orbits as their shapes are different; $r_{b1}$ and $r_{b2}$ are the coefficients of the exponential short-term ramp for the other orbits; $R_p / R_S $ is the radius ratio;
 $T_{mid}$ is the mid-transit time.\\

We obtained the different radius ratio for each wavelength, which gives the transmission spectrum.\\

\begin{figure}[h!] 
    \centering
    \includegraphics[width=0.5\textwidth]{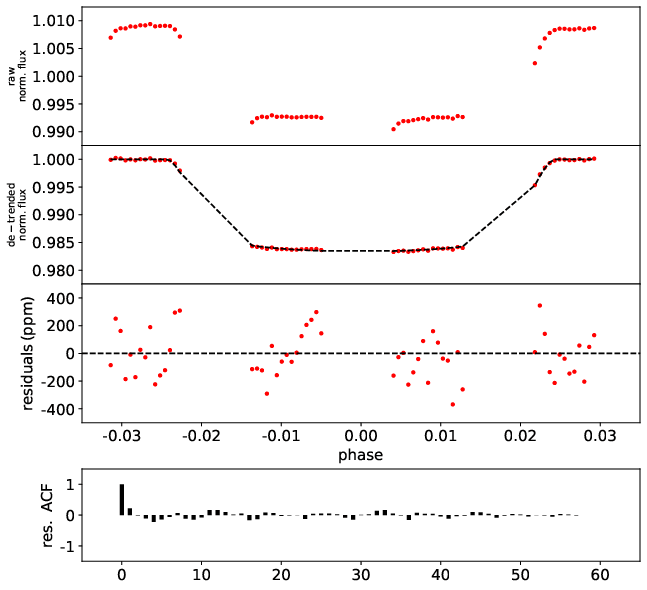}
    \caption{White light curve analysis for the G141 grism observation of WASP-17b taken on 2012/07/23. From top to bottom : Normalised raw light curve, light curve
divided by the best-fit model for the systematics, fitting residuals, and auto-correlation function of residuals. }
    \label{fig:g141_white_fitting_w17b}
\end{figure}

\begin{figure}[h!]
    \centering
    \includegraphics[width=0.5\textwidth]{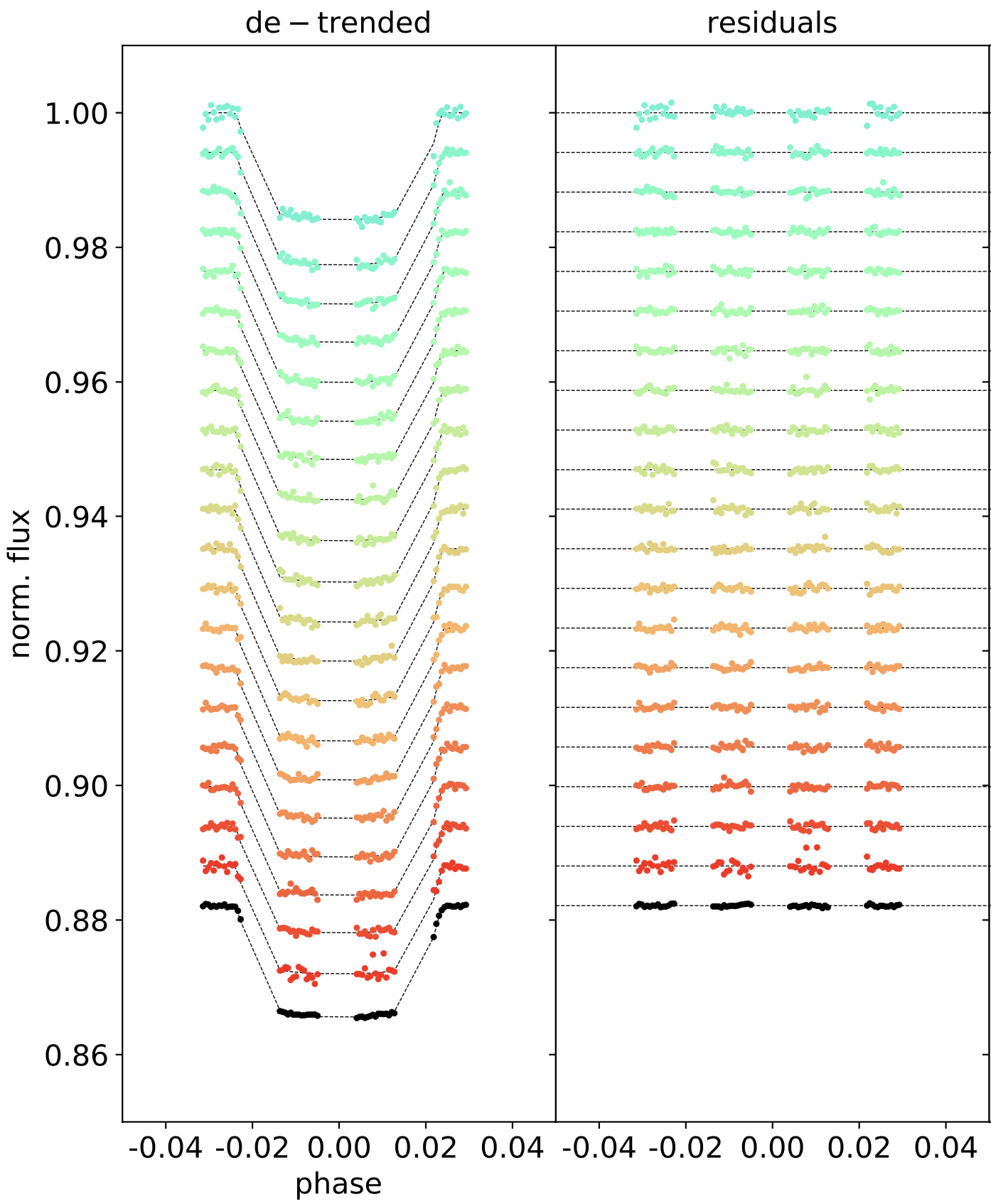}
    \caption{Spectral light curve analysis for the G141 grism observation of WASP-17b taken on 2012/07/23. Left panel: White light curve in black, spectral light curve at 1.66 µm in red, and spectral light curve at 1.1 µm in blue. Right panel: Corresponding residuals.}
    \label{fig:all_fitting_w17b}
\end{figure}

\begin{figure}[h!]
    \centering
    \includegraphics[width=0.5\textwidth]{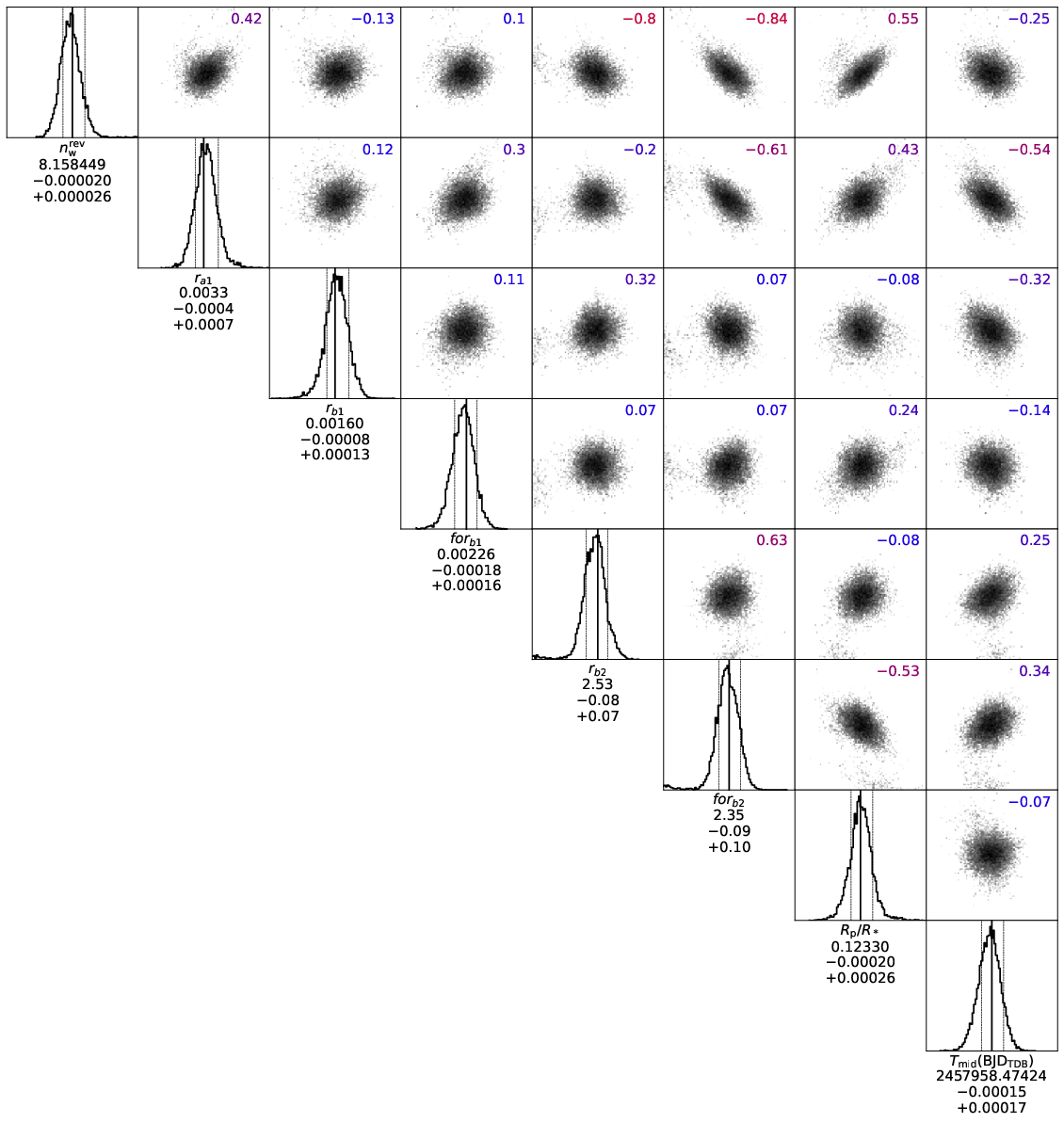}
    \caption{Correlations between the fitted parameters of the transit-shape model for the white light curve of the G141 grism observation of WASP-17b taken on 2012/07/23. }
    \label{fig:white_correlations_w17b}
\end{figure}

\subsection{Comparisons with previous works}
\label{section:2-3comp}

We compared our reductions with data described in the literature to check whether they are consistent. For planets for which there is WFC3 data, we compared with \cite{Sing_2015}. In addition, we also made comparisons with \cite{Line_2013}, \cite{Wong} (HAT-P-12b), \cite{Carter2020} (WASP-6b), \cite{Saba}, \cite{Alderson} (WASP-17b), and \cite{Wakeford_2017} (WASP-39b). We obtained small offsets that can be explained by the differences in the reduction process. In the Iraclis pipeline, a sky-background subtraction step was added. It was also demonstrated that such offsets came from a difference in treatment in long term baseline \citep{Guo,Yip_2020, Changeat_2022}. Cosmic ray and bad pixel correction are also calculated differently in Iraclis.\\

For certain planets (WASP-17b, WASP-39b, HAT-P-12b), we had several visits and in consequence several data sets. In those cases, we calculated a mean for our different radius ratio to have more signal and more data.\\

We still observed an offset between the data sets of different instruments on WASP-6b. There is a gap between the STIS data set from \cite{Sing_2015} and our own reduction of the WFC3 instrument. There is no WFC3 data for WASP-6b in the \cite{Sing_2015}, so there is no overlap in this wavelength band. To correct this inter-calibration effect between instruments, we did an interpolation on the Sing data and then removed the average difference between the two curves, the interpolation, and our spectrum. This offset is only present for WASP-6b, so we did not apply any correction for the other targets.\\

Even after these corrections, we still noted small differences on our WASP-6b reduction and the reduction presented in \cite{Carter2020}, that is, mostly a small shift in wavelength. We also see a small vertical offset on WASP-39b between studies that use Iraclis (\cite{Tsiaras_2018}, this study) and \cite{Wakeford_2017}. The effect of choosing consistent orbital parameters was discussed in \cite{Alexoudi}, especially for data with visible wavelength coverage. However, even when the exact same planetary and stellar parameters were taken into account, it was seen that the treatment of the long-term trend in the transit model create offsets between reductions \citep{Yip_2020,Changeat_2022}. This long-term trend can be describe as linear, quadratic, exponential, and so on. \cite{Guo} experimented with various different types of trends and showed that it has an impact on the transit depth retrieved from the white light curve.

To avoid inducing errors due to the use of data from several instruments, the analysis detailed in the following parts was carried out on three different data sets: a complete data set, a data set from STIS, and a data set from WFC3 for each planet.
This allows us to check the consistency of our results.\\ 

\begin{figure*}
   \centering
   \includegraphics[width=\textwidth]{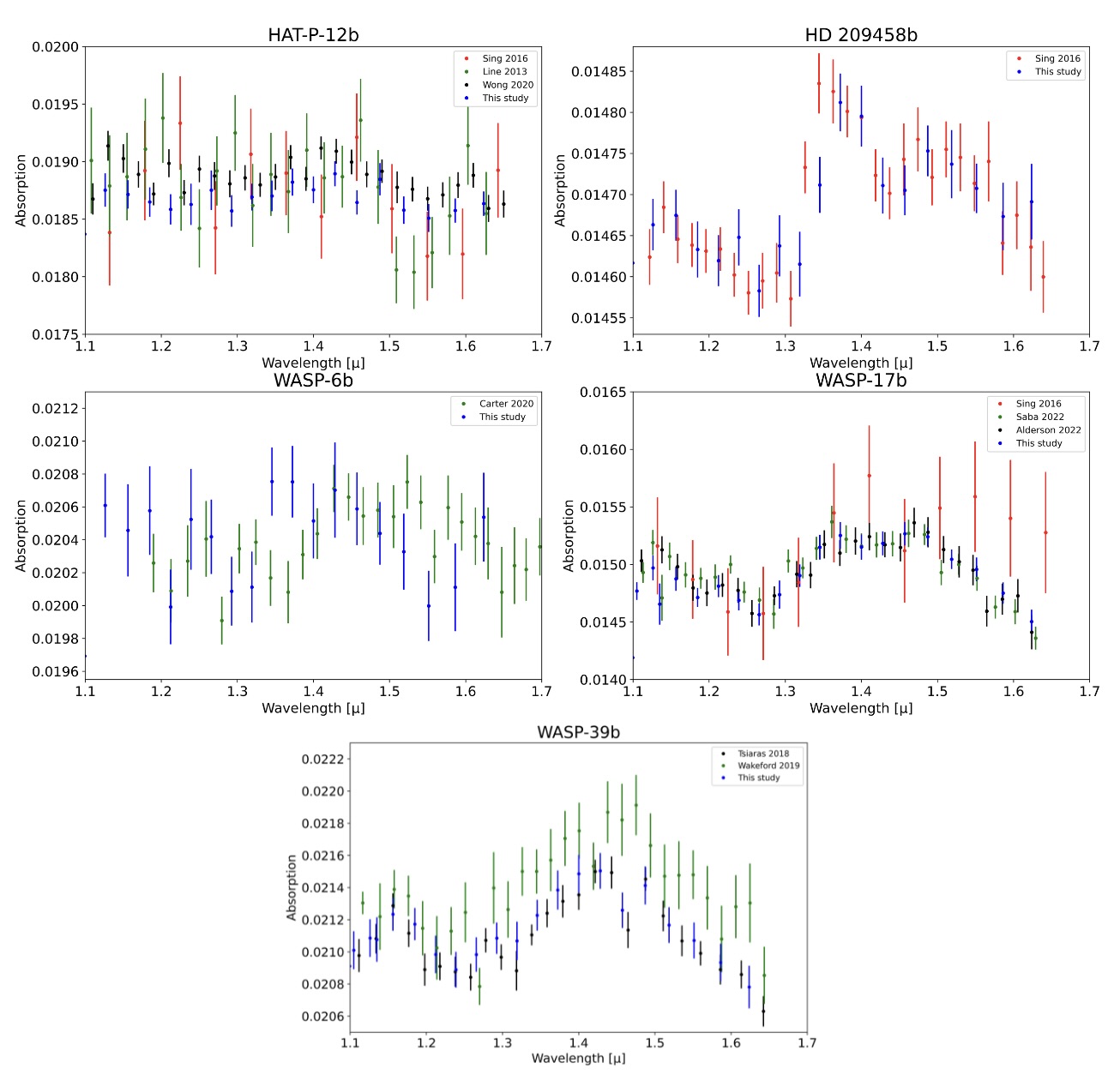}
   \caption{Comparison of data sets between our reduced data and earlier studies. The reduced data sets presented here are represented by the blue dots in each figure.}
              \label{figure_comparaison}
\end{figure*}

Once we had our data sets complete, we used the TauREx program to help retrieve the characteristics of the atmosphere. The revisit of the previous HST observations of \cite{Sing_2015} is justified by the addition of retrieval in the method, which is a statistical approach that allows us to make fewer assumptions than forward modeling (described below).
\\

{\renewcommand{\arraystretch}{1.4}
\begin{table*}
\caption{Results of the reduction re-analysis: Transit depths and errorbars for our five targets.}
\label{tab:params_depths}
\centering
\begin{tabular}{cccccc}
\hline\hline
$\lambda$ & HAT-P-12b & HD209458 b & WASP-6b & WASP-17b & WASP-39b\\
$\mu m$ & $(R_p/R_*)^2$ [\%]&$(R_p/R_*)^2$ [\%] & $(R_p/R_*)^2$ [\%] & $(R_p/R_*)^2$ [\%] & $(R_p/R_*)^2$ [\%] \\\hline
1.0994 & 1.837$\pm$0.022 & 1.462$\pm$0.004 & 1.969$\pm$0.037&1.419$\pm$0.020 &2.091$\pm$0.021 \\
1.1262 &1.875$\pm$0.015 & 1.466$\pm$0.003 & 2.061$\pm$0.019& 1.497$\pm$0.011 &2.109$\pm$0.012\\
1.15625& 1.872$\pm$0.012 &1.467$\pm$0.003 & 2.046$\pm$0.028 & 1.488$\pm$0.010 &2.123$\pm$0.010 \\
1.18485& 1.865$\pm$0.013 & 1.463$\pm$0.003 & 2.058$\pm$0.027 & 1.471$\pm$0.008 &2.117$\pm$0.010 \\
1.21225& 1.859$\pm$0.013 & 1.462$\pm$0.003 & 1.999$\pm$0.023 & 1.481$\pm$0.010 & 2.098$\pm$0.012 \\
1.23895& 1.863$\pm$0.018 & 1.465$\pm$0.003 & 2.052$\pm$0.031 & 1.469$\pm$0.009 & 2.089$\pm$0.011\\
1.26565 &1.875$\pm$0.017 & 1.458$\pm$0.003 & 2.042$\pm$0.023 & 1.456$\pm$0.010 & 2.098$\pm$0.011\\
1.29245 &1.857$\pm$0.014 & 1.464$\pm$0.004 & 2.009$\pm$0.021 &1.474$\pm$0.012 & 2.109$\pm$0.010\\
1.31895 &1.869$\pm$0.012 & 1.462$\pm$0.004 & 2.011$\pm$0.022 &1.490$\pm$0.010 & 2.107$\pm$0.0120 \\
1.34535 &1.870$\pm$0.013 & 1.471$\pm$0.003 & 2.075$\pm$0.021 &1.515$\pm$0.010 &2.123$\pm$0.010 \\
1.3723 &1.882$\pm$0.012 & 1.481$\pm$0.004 & 2.075$\pm$0.022 & 1.525$\pm$0.012 &2.138$\pm$0.012 \\
1.4 &1.876$\pm$0.012 & 1.480$\pm$0.004 & 2.051$\pm$0.023 & 1.515$\pm$0.011&2.149$\pm$0.012 \\
1.42825 &1.890$\pm$0.011 & 1.471$\pm$0.003 & 2.070$\pm$0.029 &1.518$\pm$0.010 & 2.150$\pm$0.011 \\
1.4572& 1.865$\pm$0.010 & 1.471$\pm$0.003 & 2.059$\pm$0.022 &1.53$\pm$0.010 & 2.126$\pm$0.011\\
1.4873& 1.885$\pm$0.014 & 1.475$\pm$0.003 & 2.044$\pm$0.019&1.524$\pm$0.008 & 2.141$\pm$0.012  \\
1.5186 &1.858$\pm$0.012 & 1.474$\pm$0.004 & 2.033$\pm$0.023 &1.505$\pm$0.008 & 2.117$\pm$0.011\\
1.55135 &1.851$\pm$0.012 & 1.471$\pm$0.003 &1.999$\pm$0.021 &1.496$\pm$0.010 &2.107$\pm$0.011 \\
1.5862 &1.856$\pm$0.010 & 1.467$\pm$0.004 &2.011$\pm$0.027 &1.475$\pm$0.009 & 2.093$\pm$0.012 \\
1.6237 &1.864$\pm$0.011 & 1.469$\pm$0.005 & 2.054$\pm$0.027 & 1.450$\pm$0.010 & 2.078$\pm$0.013  \\
1.6616 &1.830$\pm$0.017& 1.472$\pm$0.008 &2.028$\pm$0.032 &1.477$\pm$0.020 & 2.099$\pm$0.016 \\
\hline \hline
\end{tabular}
\end{table*}
}


\section{Methodology}
\label{section:3method}

TauREx is a radiative transfer calculation and atmosphere modeling program developed at the University College of London (UCL) \citep{AlRefaie, Al_Refaie_chemcomp}. This program makes it possible to simulate transmission spectra according to defined parameters (composition, temperature, clouds, etc.). We can directly calculate synthetic spectra in direct models or do retrievals with a Bayesian analysis using methods such as nested sampling.\\

\subsection{TauREx set-up}
\label{subsection:taurex_setup}

For all the cases studied in this paper, we simulated atmospheres with 100 layers between $10^{-5}$ Pa and $10^7$ Pa. For retrievals, we conducted a Bayesian analysis using PyMultinest (\cite{taurex_multinest}, \cite{Buchner}) with a tolerance of 0.5 and 200 to 1000 live points (10 to 20 times the number of fitted parameters).\\

We used the stellar and planetary parameters presented in Table \ref{tab:parameters}. We used three different type of models with different chemistry computations. We first used the TauREx free chemistry set-up: we included a list of molecules based on previous literature and we retrieved one abundance constant throughout the atmosphere for each molecule. We decided to include 11 molecules in this model type : H$_2$O, CH$_4$, NH$_3$, HCN, CO, CO$_2$, Na, K, H$_2$S, TiH, and AlO. We mainly considered TiH and AlO for WASP-17b \citep{Saba}, and H$_2$S for WASP-39b \citep{jwst_w39b}, but we nonetheless included these three molecules for every target. We first used two different equilibrium chemistry codes: ACE \citep{Agundez_2012,Agundez_2020} and Fastchem \citep{Stock_fastchem}, which are described in more detail in Section \ref{section:3.2chemistry}. These three computations were applied with and without the addition of clouds in the models, on our three data sets for each of the five targets, amounting to 30 models in total. We then ran every configuration on three different types of data sets: the complete data set, the STIS data set alone and the WFC3 data set alone, for a total of 90 models. Our results on the smaller data sets are discussed in Section \ref{section:cut_datasets}.\\

Regarding the temperature profile, we decided to keep a simple isothermal profile for every target, as previous works in the literature have considered it to be a good approximation \citep{Sing_2015,Pinhas}. Several tests have been made with variable T(P) profiles, with only marginal differences in the retrievals because the pressure range probed by the observations is very narrow \citep{Rocchetto_2016}. We are aware of the bias that this can induce in our results and are currently working on a more in-depth study concerning the impact of the TP profile on our models.\\

\subsection{Using equilibrium chemistry}
\label{section:3.2chemistry}

We then added equilibrium chemistry to our transmission models, using the ACE code available in TauREx.
The ACE code has already been explained in detail in several papers \citep{Agundez_2012, Agundez_2020,Al_Refaie_chemcomp,alrefaie2022freckll}, so we give a brief summary of the method used below.\\

For a closed system of N chemical compounds at a certain temperature and pressure, in the absence of disturbance (transport, UV radiation for example), the chemical composition at equilibrium can be calculated theoretically, thanks to thermodynamics quantities such as entropy, enthalpy, and so on. These quantities are calculated using NASA polynomials. At equilibrium, a system will have the chemical composition that will minimize its Gibbs energy. To compute a thermodynamic equilibrium, the ACE code takes a given pressure and temperature couple. From that couple, the chemical composition that gives the lowest Gibbs energy to the system can be
calculated. \\

The ACE code is mainly focused on neutral species composed of C, H, O, and N and has been validated over a wide range of temperature-pressure conditions, up to 2500K and 100bar. It is therefore well suited to modeling hot Jupiter planets.\\

The Fastchem code ranges down to 100 K and up to 1000 bar. It uses 396 neutral and 114 charge species sourced from the NIST-JANAF database \citep{chase1986}. We decided to use this code because the ACE equilibrium chemistry does not take into account sodium and potassium (two species that we describe in more detail in Section \ref{section:3.4contrib_Na_K} ). \\

\subsection{Clouds parametrization}

Cloud and haze opacities are known to strongly affect the shape of the spectral retrievals, to the point of canceling the molecular absorptions in some cases. An optically thick deck of clouds at a certain altitude will block us from seeing any absorption underneath this altitude. As a result, it will flatten the entire spectrum. Hazes, on the other hand, will increase the scattering slope, so it will have a stronger effect at shorter wavelength. That is why we decided to take into account optically thick clouds as well as hazes.\\

We considered two parameterizations, starting with an absorbing cloud deck in the atmosphere, from which we retrieved the pressure at which the deck is located, hereafter referred to as "clouds." We also took into account hazes, physically represented by Mie scattering contribution to the optical depth formalism \citep{Lee}. We can retrieve the particle size and mixing ratio in the atmosphere,  hereafter referred to as "hazes." \\

\subsection{Contributions of Na and K}
\label{section:3.4contrib_Na_K}
Recently, \cite{Casasayas, Casasayas2, Morello_Na} questioned the detection of atomic lines in the atmospheric spectra of one of our targets, namely, HD 209458b 
at high resolution, due to stellar interference with planetary lines in the Rossiter-McLaughlin effect. The effect has been observed for the detection of sodium lines, but also for potassium lines. This effect would impact not only HD209458b, but also every planet in our sample.\\

Without entering the debate on the presence of the atomic lines in exoplanet spectrum, the question addressed here will be on the impact of the presence (or not) of alkali lines in the exoplanet spectrum, on the retrieval of other molecular absorptions, in order to develop the consequences of the presence of alkali on the spectral retrieval of other molecules.\\

Depending on how we take the Rossiter-Mclaughlin effect into account, we could observe lines of the star in the spectrum of the planet. We chose to characterize the contributions of these two atomic species to remove the observation points that are "contaminated" in some way.\\

\begin{figure*}
   \centering
   \includegraphics[width=\textwidth]{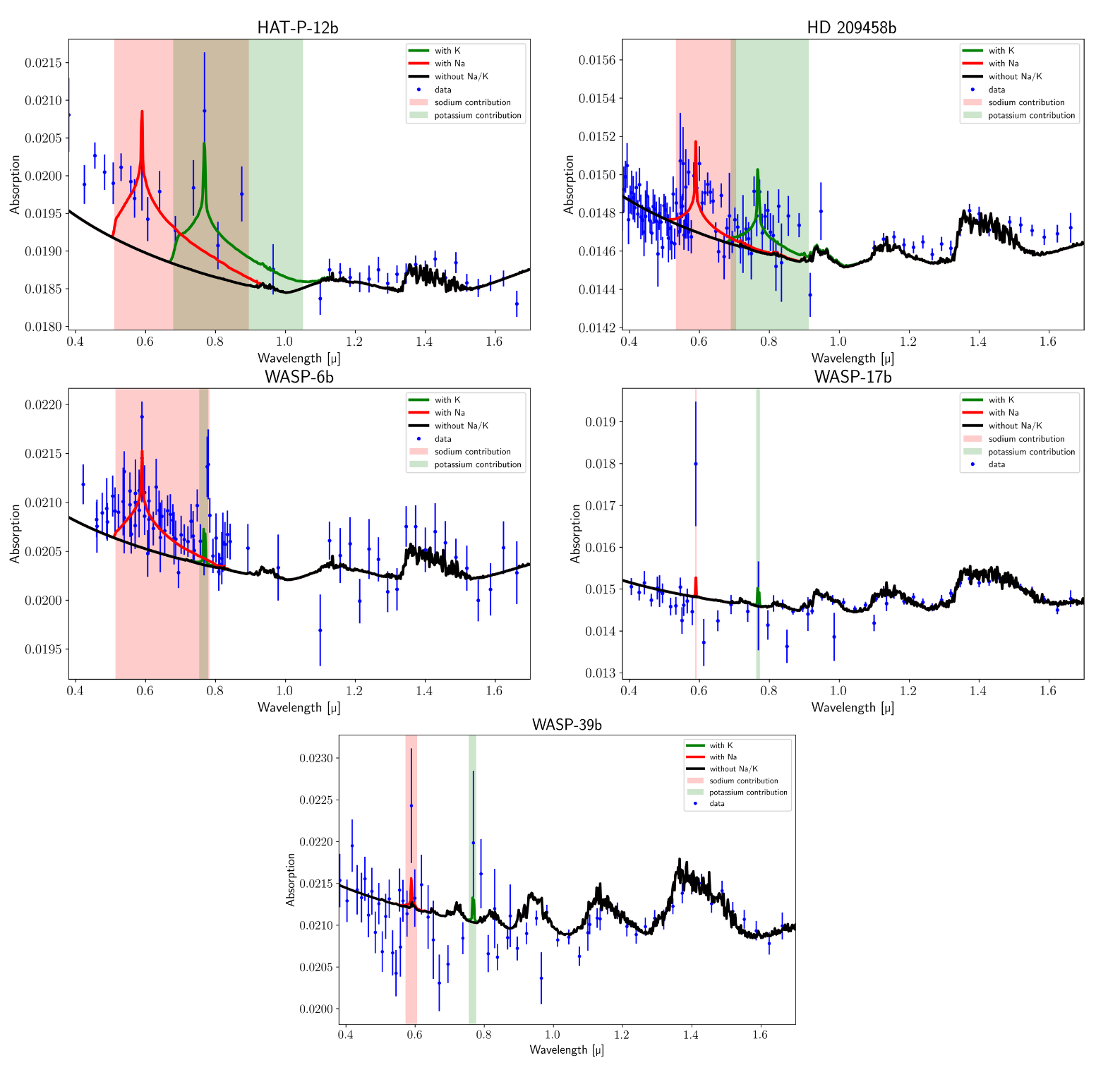}
    \caption{Comparison of direct models on our five targets. The black line is the reference model and doesn't include any sodium or potassium absorption contributions. The red line is the same model as the reference one with Na absorption added and the green line is the same with K absorption added. The red and green shadows correspond to the contributions of sodium and potassium, respectively.}
    \label{fig:na_k_contrib_5_planetes}
\end{figure*}

By comparing our data set with models of absorption contribution of sodium and potassium, we removed points that were considered to have been contaminated by these absorptions. Absorption contribution of the two species are plotted in Figure \ref{fig:na_k_contrib_5_planetes} for our targets. The line profiles have fairly large wings, so to be sure to take into account the whole contribution consistently, we removed points from about 0.5µ up to the first WFC3 data point for every target. That way, we still had the information given by the slope at these wavelengths, which offers important clues on the physics of the clouds. We then computed a retrieval analysis with two different data sets: one complete and one without the contaminated data points. We applied a similar parametrization in both cases to see whether the retrievals results would be significantly different and if the presence of Na and K absorptions would bias other retrieved parameters. We did this analysis for the five targets presented here.\\

\section{Planet-by-planet results}
\label{section:4results}

\begin{figure*}
   \centering
   \includegraphics[width=\textwidth]{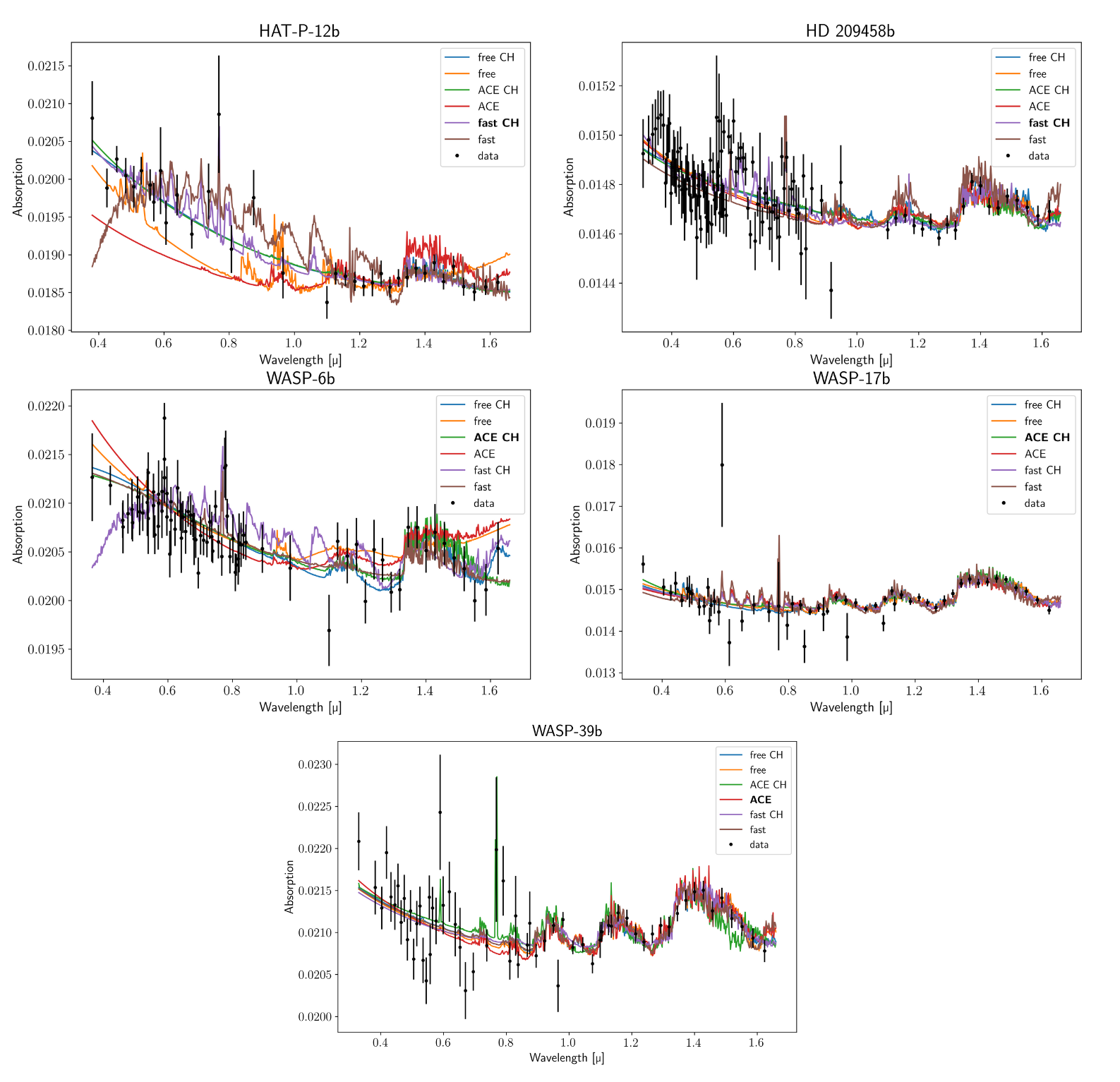}
   \caption{Model comparison for each planet. Details on parameters used and contributions that has been taken into account inside every model can be found in Section \ref{subsection:taurex_setup}. We can see that the WFC3 part of the spectra is most of the time well-fitted and we have discrepancies at shorter wavelength. Here, we use our whole data sets with Na and K contribution points.}
              
   \label{figure_comparaison}
\end{figure*}

{\renewcommand{\arraystretch}{1.4}
\begin{table*}[h!]
\caption{Bayesian factor comparison between every models performed on a full data set (HST STIS + HST WFC3) for our five planets. We defined the Bayesian factor as the difference between the minimum log(evidence) and the log(evidence) of the compared model, for a given planet. The best-fit for every planet is indicated in bold.}
\label{tab:bayesfactor}
\centering
\begin{tabular}{|c|c|c|c|c|c|c|}
\hline\hline
 & \multicolumn{2}{c|}{Free chemistry} & \multicolumn{2}{c|}{ACE} & \multicolumn{2}{c|}{Fastchem} \\\hline 
  & no clouds & clouds & no clouds & clouds & no clouds & clouds\\\hline 
HAT-P-12b & 49.2 & 87.3 & 0 & 89.7 & 58.5 & \textbf{90.8} \\
HD 209458b & 24.3 & 28.3 & 26.6 & 33.1 &0&\textbf{44.2}\\
WASP-6b & 23.5 & 35.9 & 16 & \textbf{38.6} & 0&32.7\\
WASP-17b &15.4&13.8&8.6&\textbf{15.8}&0&10.4\\
WASP-39b & 4.6& 2 &\textbf{10.2}&0&5.3&5.3\\
\hline\hline 
\end{tabular}
\end{table*}
}

After seeing the data reduction process and the method used, we go on to discuss the results of the study. These results are presented here planet by planet, with an additional section on the results concerning the effect of sodium and potassium contributions, which concern all five planets.\\

To avoid any bias due to the assembling of data from several instruments, we decided to analyze the data from the two instruments separately and then together, namely, using three sets of data per planet.
However, the data from the short wavelength STIS instrument presents fewer spectral signatures and will therefore be less reliable for the found abundances of the molecules. As for the WFC3 data, it will be less reliable in determining the presence or absence of clouds in the atmosphere because of the wavelength coverage starting further away. This is why setting the two instruments in tandem is an interesting step.\\

Different notations will be used to refer to the three model types. "Free" uses the taurex chemistry parametrization, namely, for each integrated molecule a constant abundance profile along a column of atmosphere. Models denoted "ACE" correspond to chemical equilibrium models that use the ACE code. Models denoted "Fchem" correspond to chemical equilibrium models that use the Fastchem code. When "CH" is indicated, it means that clouds and hazes have been added to the model. Only "H" means that only hazes have been added and "C" indicates that only clouds have been added.\\

We made a comparison table with the Bayesian factor, set as Table \ref{tab:bayesfactor}. We defined here the Bayes factor as the difference between the minimum log(evidence) and the log(evidence) of the compared model, for a given planet. This allows us to compare the differences between several models in a consistent way, regardless of the target mentioned.\\

\subsection{Cut data sets}
\label{section:cut_datasets}

We based each of the six models on HST STIS data sets alone and on HST WFC3 data sets alone to check that we did not have any bias due to the assembling of data from several instruments. It is much more difficult to conclude on these models: the log(evidence) are always quite similar between the models. The STIS data nevertheless show us a small increase in log(evidence) for the consideration with clouds, consistent for the three chemistry set-ups on the planets HAT-P-12b and HD 209458b. This result is consistent with the conclusions made in the following parts (Sections \ref{section:4.2hp12_results} and \ref{section:4.3hd209_results}). It is impossible to distinguish between the clear atmosphere hypothesis and the cloudy atmosphere hypothesis on the other three planets (WASP-6b, WASP-17b, and WASP-39b), which suggests a clear atmosphere.\\

On the WFC3 data alone, the log(evidence) is very similar, so this does not allow us to conclude on the presence or absence of clouds in the atmosphere of our targets, as expected. Nevertheless all best-fit models include the ACE equilibrium chemistry. This is quite consistent with the results presented on the entire data sets (see Table \ref{tab:bayesfactor}), but on the complete data, the best-fit fluctuates between ACE equilibrium chemistry and Fastchem equilibrium chemistry.\\

\begin{figure}[h!]
    \centering
    \includegraphics[width=0.5\textwidth]{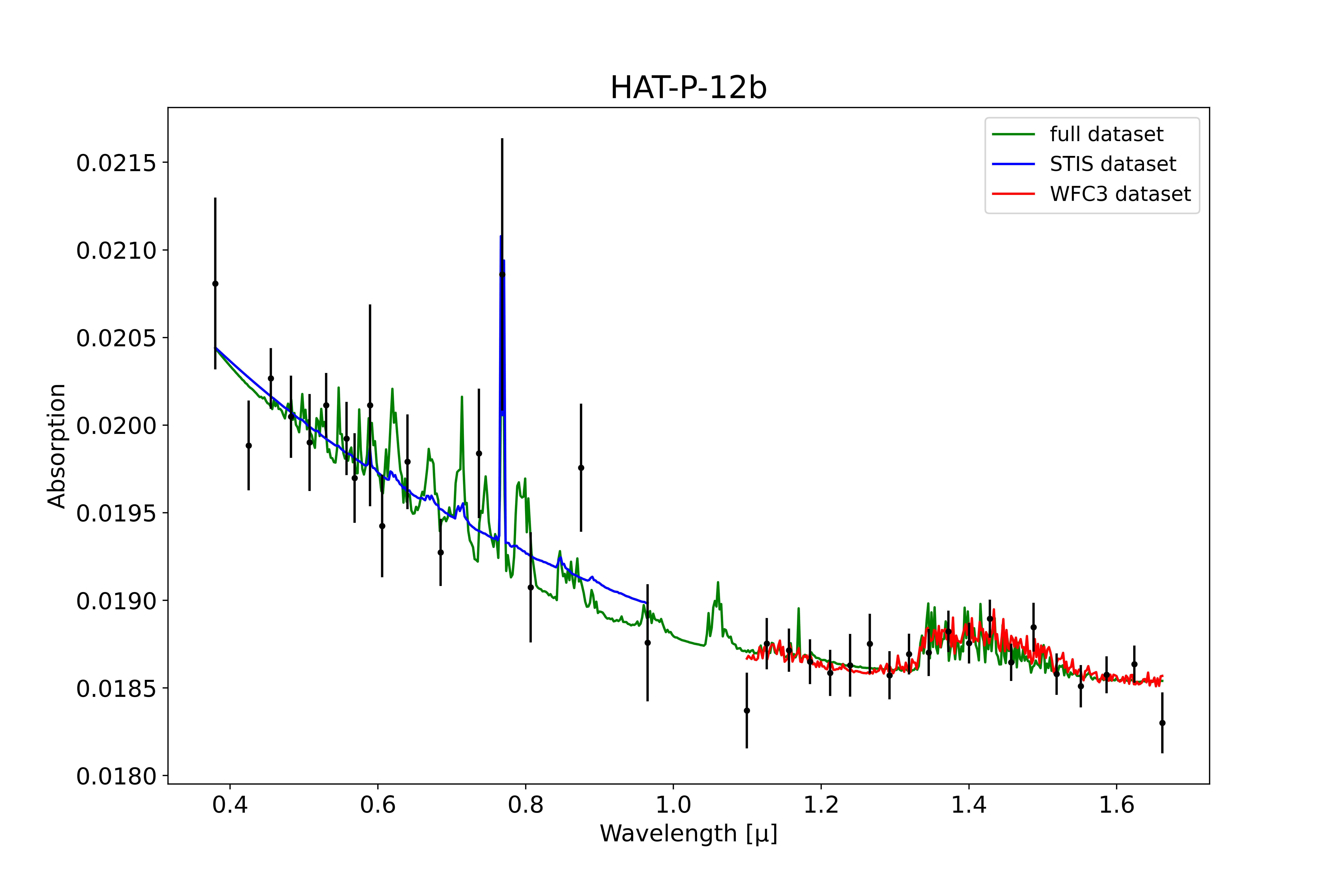}
    \caption{Models retrieved on the full data set (green curve), the STIS data set alone (blue curve), and the WFC3 data set alone (red curve). This figure shows the differences between the spectra for HAT-P-12b best-fit model and the FastChem with cloud consideration. }
    \label{fig:hp_3datasets}
\end{figure}

For example, in Figure \ref{fig:hp_3datasets}, we can see the differences between the models that has been retrieved on the three different types of data sets. We can see that models computed on the complete and WFC3 data sets are very similar, but the differences between the model with and without clouds are minimal and the log(evidence) are really close. We cannot draw a conclusion on the absence or presence of clouds, however, we can conclude on the clouds on the model run on the full data set, thanks to the addition of the STIS data. We can also see that we have more optical absorbers and less water on the model calculated on STIS data set alone, which was expected due to the wavelength coverage and the position of the absorption features. Overall, the three models are consistent with one another and in agreement with our conclusions, developed in Section \ref{section:6:conclusion}. 

\subsection{HAT-P-12b}
\label{section:4.2hp12_results}

For HAT-P-12b, \cite{Sing_2015} showed a strong optical slope due to Rayleigh scattering, as well as aerosols hazes and a potassium feature. \cite{Deibert} also found a sodium absorption feature.
However, \cite{Wong} and \cite{Yan} did not find any detected alkali absorption peaks. They still agreed on the cloudy, hazy atmosphere of the planet with a prominent Rayleigh scattering slope on the visible wavelengths.
\cite{Jiang} also showed no alkali absorption signatures, but did not find any evidence of a cloudy or hazy atmosphere. They additionally explained the discrepancies of the transmission spectra by the effect of stellar activity as unocculted spots and faculae. Based on these previous results, we decided to try several models to fit our data. We explored three different chemistry assumptions. The first "free" model that takes into account absorption contributions of H$_2$O, CH$_4$, NH$_3$, HCN, CO, CO$_2$, Na, K, H$_2$S, TiH, and AlO. The second chemistry type is the "ACE" model and the third one uses the Fastchem code. We tested each one with or without clouds or hazes.

For this planet, we can see clear differences between models that include clouds and hazes and models that do not include them. This 
 allows us to conclude that the atmosphere 
of HAT-P-12b is probably cloudy. We cannot however statistically distinguish between our three chemistry parametrization, as their log(evidence) is similar. We note that visually the Fastchem models could over-fit the HST STIS part of the spectrum.
Overall, the model with the most significant log(evidence) is the Fastchem model with clouds or hazes (log(evidence)=252.7). The posterior distribution figure for the best-fit model is given in Appendix \ref{section:appendix_hp12}.

\subsection{HD 209458b}
\label{section:4.3hd209_results}

\cite{Sing_2015} found signatures of Na, H$_2$O and aerosols hazes in the atmosphere of HD 209458b.
\cite{MacDonald} detected NH$_3$ absorption feature and suggested disequilibrium chemistry processes such as vertical mixing or photochemistry.

\cite{Hawker} confirmed detections of H$_2$O and CO in the atmosphere of HD 209458b and reported some evidence for HCN.
\cite{Pinhas} concluded that this planet has hazes or clouds in its atmosphere and weak evidence for stellar heterogeneity.
This planet has been extensively studied with atmospheric kinetic models \citep{Venot_2012,Venot_2020,Moses_2011,Tsai_2017,Drummond}. All agreed on the main chemical compounds, which are H$_2$O, CO, CH$_4$, and NH$_3$, even if the predicted mixing ratios varies depending on models and chemical schemes \citep{Moses_2014,Venot_2020}. We then decided on putting H$_2$O, CH$_4$, NH$_3$, HCN, CO, CO$_2$, Na, K, H$_2$S, TiH, and AlO for the free model. We also used two equilibrium chemistry parametrizations with ACE and Fastchem and carried out two sets of runs with or without clouds and hazes.

The two Fastchem models are the best fit (with clouds) and the worst fit (without clouds), with a strong difference between the two. This clear difference is not found in the other two different chemistry set-up, with a difference between cloudy atmosphere and clear atmosphere of less than 10 in terms of the Bayes factor.
All four models have very close log(evidence) for this planet and it is very hard to untangle them. The best log(evidence) is for the model with Fastchem equilibrium chemistry and clouds and hazes (log(evidence) = 860.3). The posterior distribution figure for the best-fit model is given in Appendix \ref{section:appendix_hd209}.

\subsection{WASP-6b}

Regarding WASP-6b, \cite{Jordan} and \cite{Nikolov} agreed with a spectrum mostly featureless but with aerosols hazes. \cite{Sing_2015} found aerosols hazes as well as a potassium absorption feature. \cite{Pinhas} pointed out evidence for the presence of stellar heterogeneities as cool spots and hot faculae and hazes. Finally, the best-fit model from \cite{Carter2020} is composed of hazes and stellar heterogeneities as well as water, sodium, and potassium features. 

First, we tried a free model taking into account H$_2$O, CH$_4$, NH$_3$, HCN, CO, CO$_2$, Na, K, H$_2$S, TiH, and AlO. We also did a run with equilibrium chemistry ACE and Fastchem, adding only hazes and no clouds to all three chemistry assumptions.

We can see in the differences of log(evidence), shown in Table \ref{tab:bayesfactor}, that for the two parametrizations in chemical equilibrium, the hazes improves the fit. This result is less obvious for free chemistry models. As was the case for HAT-P-12b, we note that the Fastchem models could over-fit the part of the spectrum corresponding to HST STIS data. In the end, the model with the best log(evidence) is the one with ACE equilibrium chemistry and hazes (log(evidence) = 563.1). The posterior distribution figure for the best-fit model is given in Appendix \ref{section:appendix_w6}.

\subsection{WASP-17b}

\cite{Sing_2015} concluded with sodium and water signatures as well as a clear atmosphere without hazes. \cite{Sedaghati} agreed and also detected a potassium signature. 
\cite{Saba} found strong H$_2$O, TiH, and AlO absorption signatures, yet without completely refuting previous detections of Na and K. \cite{Alderson}, on the other hand, reported an H$_2$O detection, as well as evidence for CO$_2$, but no sodium nor potassium absorption.
\cite{Pinhas} showed that the model that best fit to his data has no stellar heterogeneity and no evidence for clouds or hazes. 

We decided to include the following species in our free model : H$_2$O, CH$_4$, NH$_3$, HCN, CO, CO$_2$, Na, K, H$_2$S, TiH, and AlO. We also ran two equilibrium chemistry models and added clouds and hazes to both models.

We obtain pretty similar log(evidence) for all six models between cloudy or clear atmosphere, which rather leans towards the hypothesis of the atmosphere without clouds or hazes for WASP-17b. We can see in Figure \ref{figure_comparaison} that all models are very similar. We note a really close log(evidence) between the free chemistry no clouds model and the ACE chemistry and clouds model, two very different models. In the end the best-fit with the highest log(evidence) is the ACE equilibrium chemistry with clouds (log(evidence) = 393.1). The posterior distribution figure for the best-fit model is given in Appendix \ref{section:appendix_w17}.

{\renewcommand{\arraystretch}{1.4}
\begin{table*}
\caption{TauREx results for the models with the highest log(evidence) for each target. We have either metallicity and C/O ratio or relative abundances depending of the model types of the best-fit.}
\label{tab:results_chemistry}
\centering
\begin{tabular}{ccccccc}
\hline\hline
Fitted parameter & bounds & HAT-P-12b & HD209458b & WASP-6b & WASP-17b & WASP39b\\ \hline
$R_p$ [R$_{jup}$]& [0.5R, 1.5R] & 1.36 &0.82 &1.07 &1.71 & 1.36\\
$T$ [K]& [500,2500] & 1 890 &1 264 & 1 110&1 350 &555\\
log(P$_{clouds}$) & [-5;7] &2.7 &2&- & 3.7& -\\
log(R$_{mie}$)& [1;-2] &-1.5 &-1.4 &-0.9 &-1.6 &- \\
log($\chi_{mie}$) & [-3;-20] & -9.6 &-10.8 & -15.6 & -17.1 &-\\
log(C/O) & [1;0] & 0.6&0.6 &-0.9 & -1.1& -1.8\\
log(metallicity) & [3;-1] &2.26 &-0.2 &-1.2 & -1.2&-0.7 \\
\hline\hline\\
\end{tabular}
\end{table*}
}

\subsection{WASP-39b}

For this planet, \cite{Sing_2015} and \cite{Fischer} found signatures of sodium and potassium without detection of aerosols hazes. 
\cite{Nikolov_w39b} agreed for the clear atmosphere with ground-based observations and detected sodium features and evidences for potassium.
However, \cite{Barstow} concluded that grey clouds or Rayleigh scattering atmosphere models were working on this planet, with a slight better fit for the grey clouds solution. 
 \cite{Wakeford_2017} found that WASP-39b is best described with isothermal equilibrium model, and found that uniform clouds were not playing an important role. 
However, \cite{Pinhas} took into account stellar activity and found the spectra best explained with stellar heterogeneities and clouds and hazes. 
\cite{Kirk} is in agreement with \cite{Wakeford_2017} and points out that stellar activity has a minimum impact on the transmission spectrum. \cite{Kawashima} recently found that WASP-39b is best described taking into account disequilibrium effect of vertical mixing for some chemical species such as H$_2$O, CH$_4$, CO, CO$_2$, NH$_3$, and N$_2$.\\

We decided to include in our free model H$_2$O, CH$_4$, NH$_3$, HCN, CO, CO$_2$, Na, K, H$_2$S, TiH and AlO and to run equilibrium chemistry ACE and Fastchem models as well. We tested with clouds and hazes added to all three models.
We added H$_2$S in our free model in agreement with \cite{jwst_w39b}, early release of JWST results on WASP-39b. 

The resulting log(evidence) for all of the models are really similar, with the clear atmosphere hypothesis therefore shown to be favored.
There is a small preference for the model with ACE equilibrium chemistry taking into account neither clouds nor hazes (log(evidence) = 443). The posterior distribution figure for the best-fit model is given in Appendix \ref{section:appendix_w39}.

\subsection{Contributions of Na and K}

The results of our reanalysis of the planets taking into account the uncertainties on the Na/K absorption features in the planetary spectra show that removing the spectral range of these features does not affect significantly the molecular retrievals of the other abundances or other parameters. Therefore, our conclusion is that there is no spurious interference effects to be expected from these absorption lines that are still under reanalysis.

It is reassuring to know that even if these detections of alkaline species turn out to be false, the other results of the studies which have taken them into account will still be correct. This result is verified for all our targets. Similar studies on other planets can be conducted as verification, but this is already a strong indication.


\section{Discussion}
\label{section:5discussion}

\subsection{Disequilibrium chemistry}

\begin{figure}[h!]
    \centering
    \includegraphics[width=0.5\textwidth]{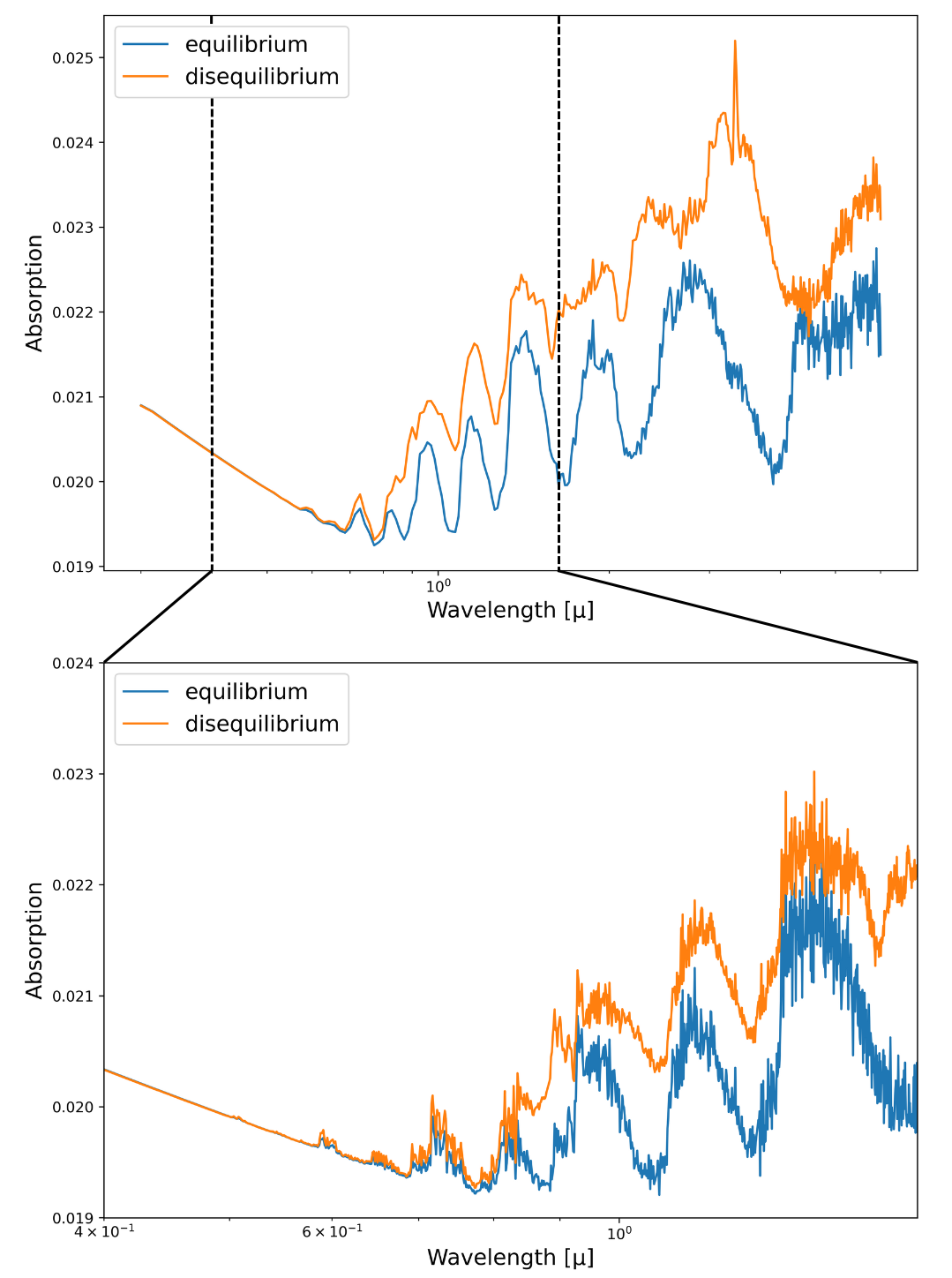}
    \caption{Forward models for HD 209458b, using equilibrium chemistry ACE (blue curve) and disequilibrium chemistry (orange curve). Bottom panel shows the effect of using those two different chemical schemes on the wavelength coverage of the data set presented in this study. The top panel shows the same models on a wider wavelength coverage.}
    \label{fig:hd_diseq_spectra}
\end{figure}

\begin{figure}[h!]
    \centering
    \includegraphics[width=0.5\textwidth]{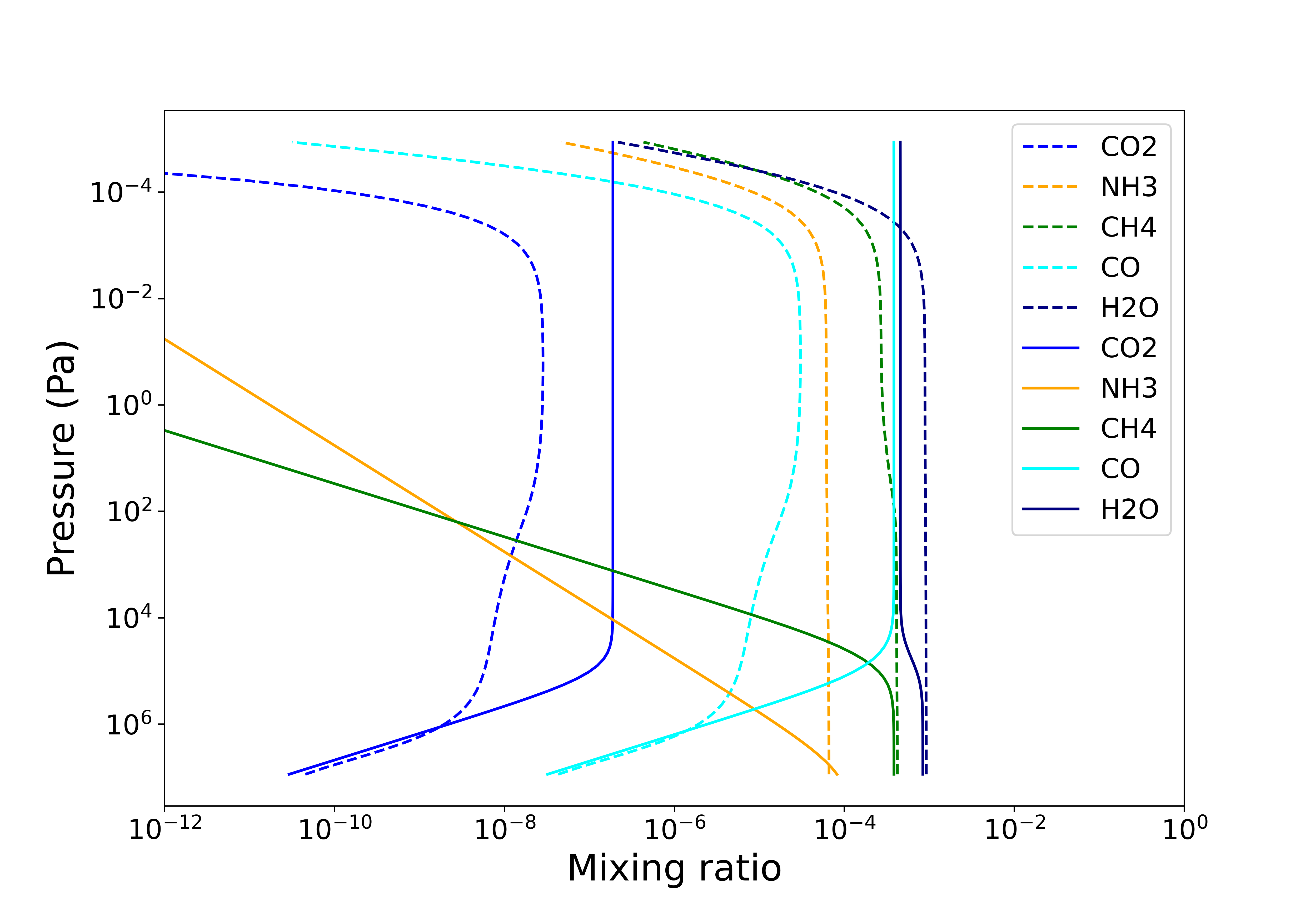}
    \caption{Abundances profiles for active molecules for the two forward models on HD 209458b comparing equilibrium chemistry (full lines) and disequilibrium chemistry (dashed lines). }
    \label{fig:hd_diseq_mixratios}
\end{figure}

We ran forward models to quantify the differences between equilibrium chemistry and disequilibrium chemistry. For this part of the study, we only used the ACE equilibrium chemistry code \citep{Agundez_2020}. For the disequilibrium chemistry, we used the FRECKLL code \citep{alrefaie2022freckll}, adapted from \cite{Venot_2020}. This section is focused on the HD209458b case with two models with the exact same planetary and stellar parameters (see Table \ref{tab:parameters}). These two models also have the same atmosphere configuration: 100 layers between 10$^{-5}$ and 10$^7$ Pa, no clouds, no hazes, isothermal TP profile, Rayleigh scattering, and collision-induced absorption between H$_2$-He and H$_2$-H$_2$. The only difference between the two models are the chemistry scheme used. The equilibrium is the blue model in Figure \ref{fig:hd_diseq_spectra} and the associated mixing ratios for active molecules are shown as full lines in Figure \ref{fig:hd_diseq_mixratios}. Disequilibrium chemistry is shown as the orange model in Figure \ref{fig:hd_diseq_spectra} and associated mixing ratios for active molecules are displayed as dashed lines in Figure \ref{fig:hd_diseq_mixratios}.\\ 

We can see the effects that taking disequilibrium chemistry into account can have on the transmission spectra, especially when we are looking at longer wavelength coverage. Any addition to the models of disequilibrium mechanisms, such as vertical mixing or photochemistry, now becomes crucial for a more realistic representation of atmospheres. The increasing amount of JWST data that will arrive in the coming years and probe more extensive parts of atmospheres makes the integration of disequilibrium processes a necessity.

\subsection{General discussion}

This study has achieved a state-of-the-art retrieval of HST observations within the frame of equilibrium chemistry. Some limitations will be addressed in future work.
First, taking into account a more complex temperature-pressure profile will be very important in future studies. Indeed, the isothermal profile remains a good first approximation because of the small wavelength coverage that we study here, meaning that the pressure range probed by the observations is very narrow. It nevertheless induces a bias in taking into account the 3D aspect of the planet. Considering a three- or four-point profile would be the next step.\\

The consideration of clouds and hazes in this study is also still quite simple, but our models are still sufficient given the precision of observational data. The study of \cite{Arfaux} looked at haziness conditions for these planets in detail.\\

Compared to a previous work, this work has pushed the limit in data retrieval of HST/WFC3 spectra with up-to-date chemical retrieval programs and allowed us to measure the composition of the five exoplanets in the hypothesis of a chemical equilibrium. Further developments will address the question of out-of-equilibrium chemistry for data retrieval. 
Having models that take as many effects as possible into consideration is an important aspect from a scientific and computational point of view. However, the key question is whether the observations will be able to follow (and with sufficient precision and at a good resolution) to allow us to fit a complex model. Nevertheless, the spectral resolution in HST/WFC3 may not allow us to distinguish between a simpler model and a more realistic model. For example, \cite{alrefaie2022freckll} introduced the chemical kinetic code FRECKLL and tested it on some simulated JWST data. Conducting some more studies on integrating disequilibrium chemistry could also be an interesting next step, especially in preparation for future JWST retrievals.


\section{Conclusion}
\label{section:6:conclusion}

In summary, this work presents an improved reanalysis of a selection of five hot Jupiters HST observations with up-to-date data reduction and advanced retrieval techniques. The list of retrieved parameters given in Table \ref{tab:results_chemistry} constitutes an update of previous results, for instance, \cite{Sing_2015}. In addition, the effect of the recent reevaluation of Na/K emission in hot Jupiters has been taken into account to show that they do not modify the infrared retrievals. This work thus constitutes a step in the HST reanalysis before the JWST spectral observations, which will open a new era in exoplanet characterization. Higher spectral resolution and higher signal-to-noise ratios from JWST will offer access to future improvements, while addressing the question of disequilibrium chemistry in retrievals \cite{alrefaie2022freckll}.


\begin{acknowledgements}
  E. Panek, J-P. Beaulieu and P. Drossart have been supported by CNES convention (6512/7493). O.V. acknowledges funding from the ANR project `EXACT' (ANR-21-CE49-0008-01), from the Centre National d'\'{E}tudes Spatiales (CNES), and from the CNRS/INSU Programme National de Plan\'etologie (PNP).   
\end{acknowledgements}

%
%

\newpage

\bibliographystyle{aa}
\bibliography{bibliotex}
\nocite{*}

\begin{appendix}
\onecolumn
\section{Posterior distributions for best-fit models}

\subsection{HAT-P12b}
\label{section:appendix_hp12}

\begin{figure*}[h!]
    \centering
    \includegraphics[width=\textwidth]{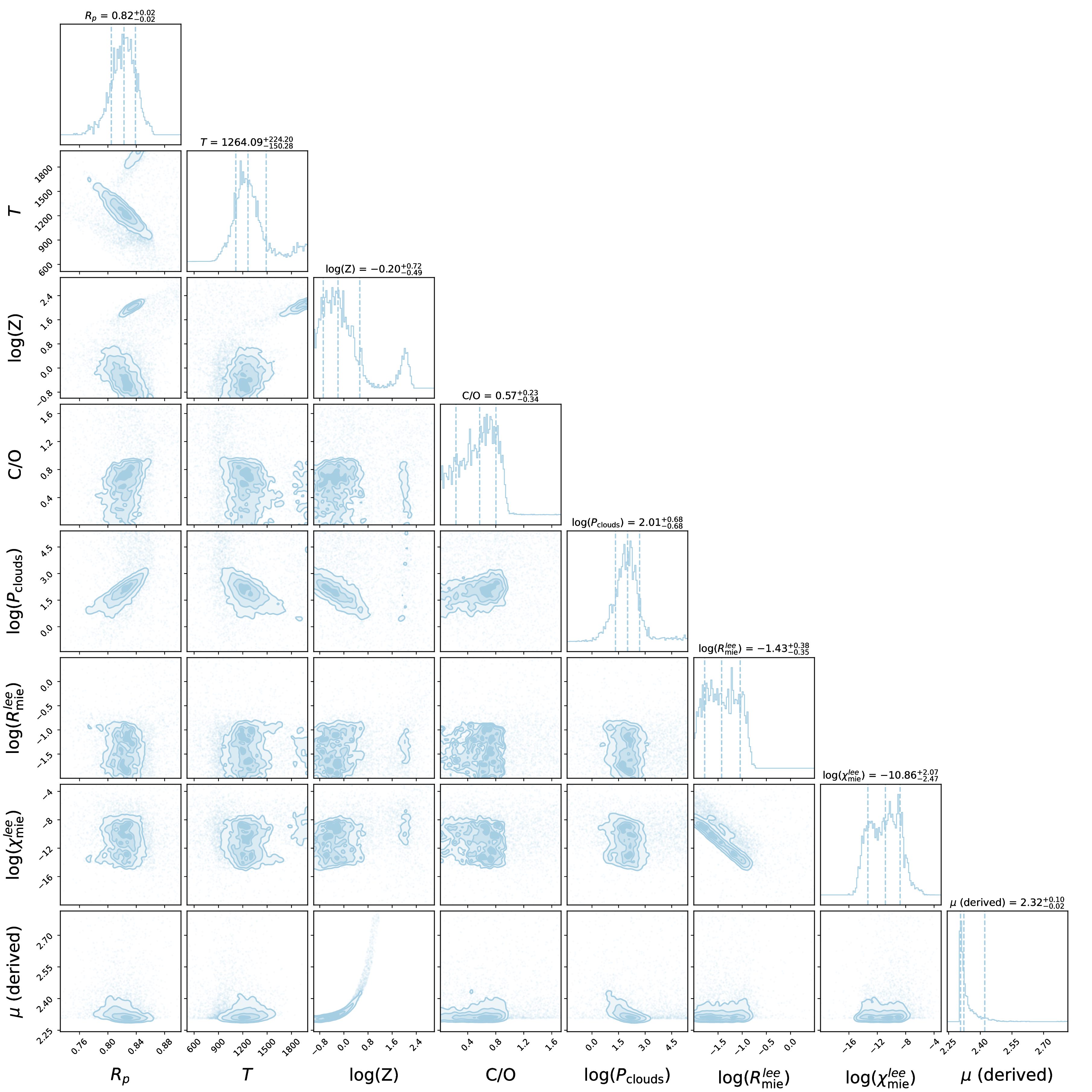}
    \caption{Posterior distribution figure for HAT-P-12b corresponding to the best-fit model with Fastchem equilibrium chemistry and clouds and hazes.}
    \label{fig:post_hp12}
\end{figure*}

\newpage
\subsection{HD 209458b}
\label{section:appendix_hd209}

\begin{figure*}[h!]
    \centering
    \includegraphics[width=\textwidth]{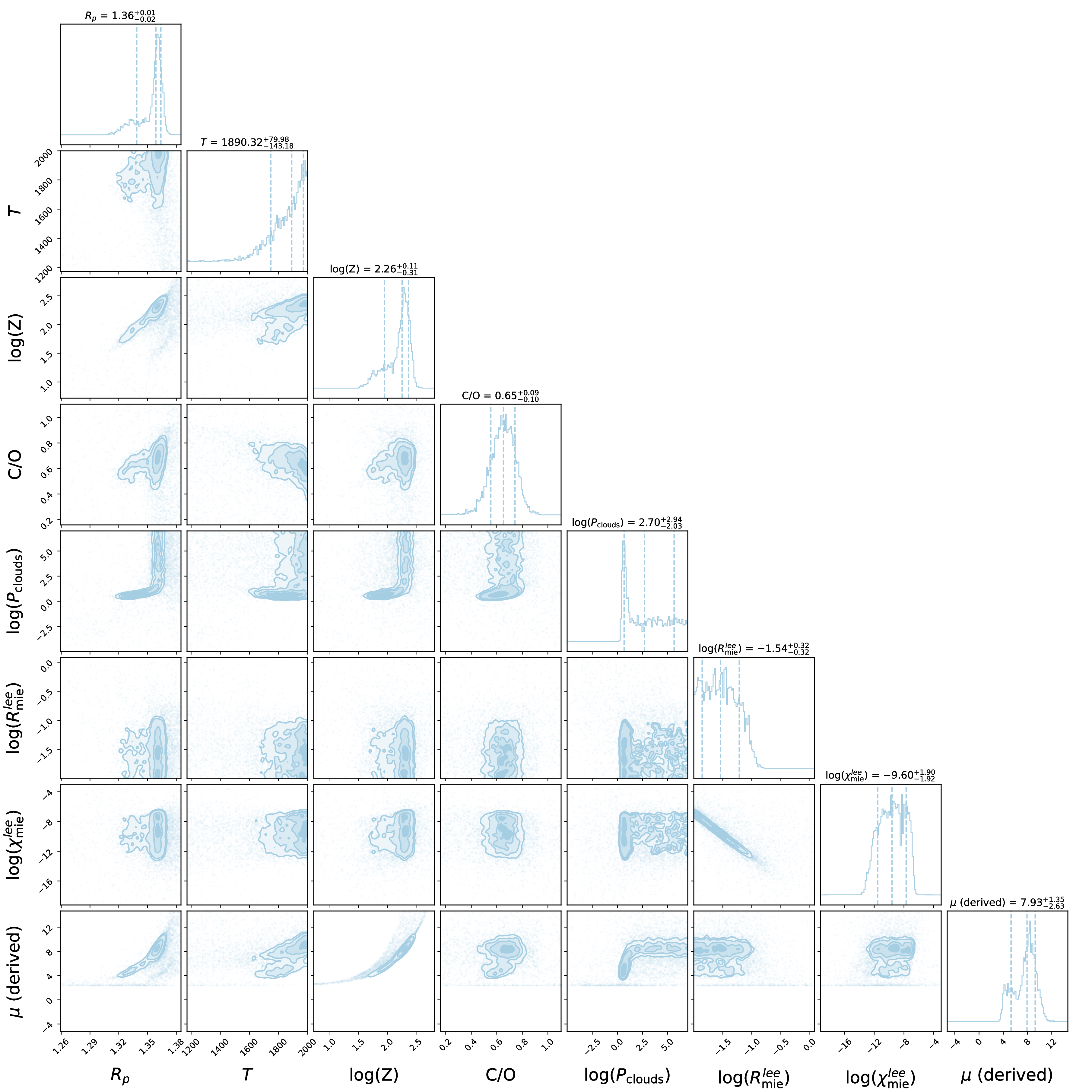}
    \caption{Posterior distribution figure for HD 209458b corresponding to the best-fit model with Fastchem equilibrium chemistry and clouds and hazes.}
    \label{fig:post_hd209}
\end{figure*}

\newpage
\subsection{WASP-6b}
\label{section:appendix_w6}

\begin{figure*}[h!]
    \centering
    \includegraphics[width=\textwidth]{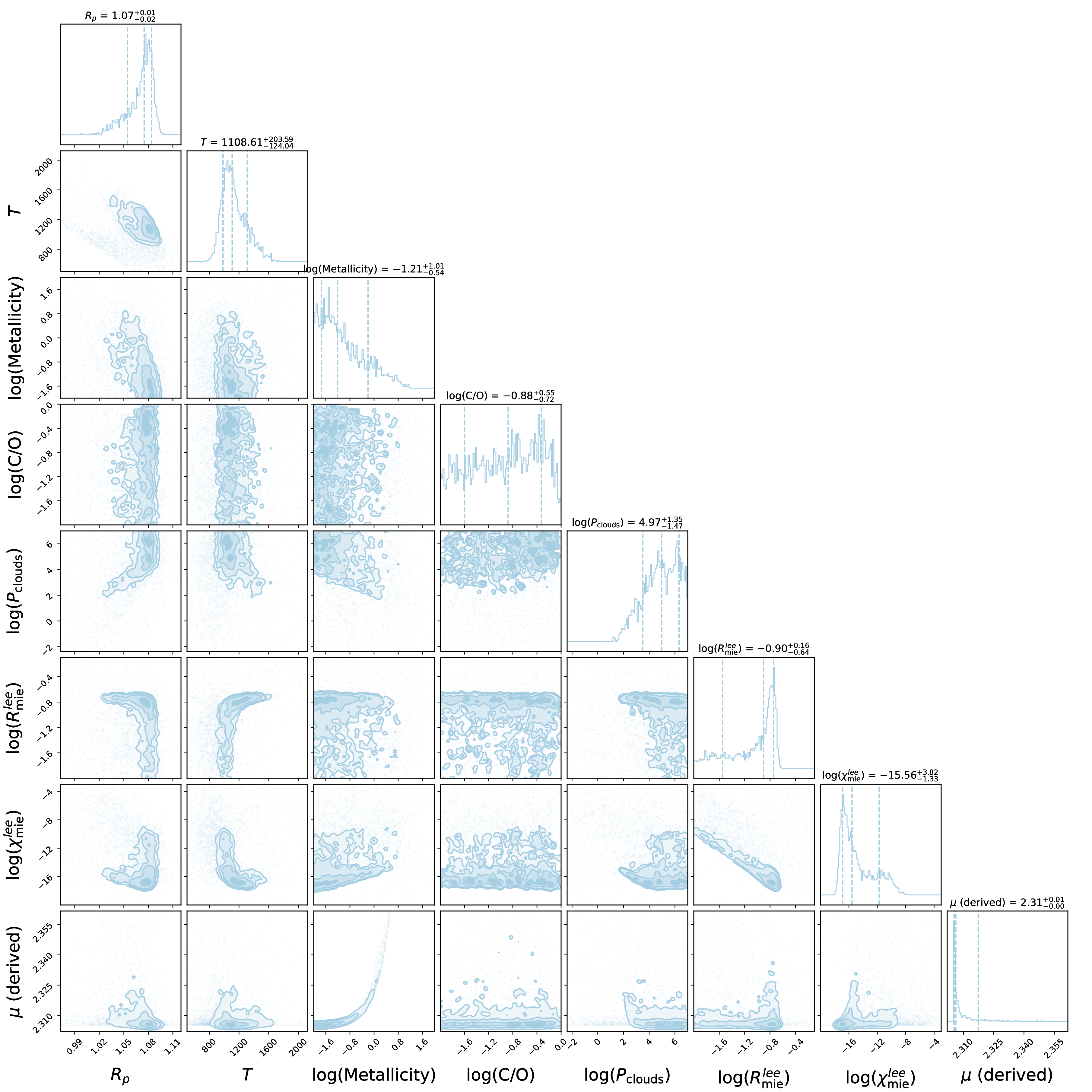}
    \caption{Posterior distribution figure for WASP-6b corresponding to the best-fit model with ACE equilibrium chemistry and hazes.}
    \label{fig:post_w6}
\end{figure*}

\newpage
\subsection{WASP-17b}
\label{section:appendix_w17}

\begin{figure*}[h!]
    \centering
    \includegraphics[width=\textwidth]{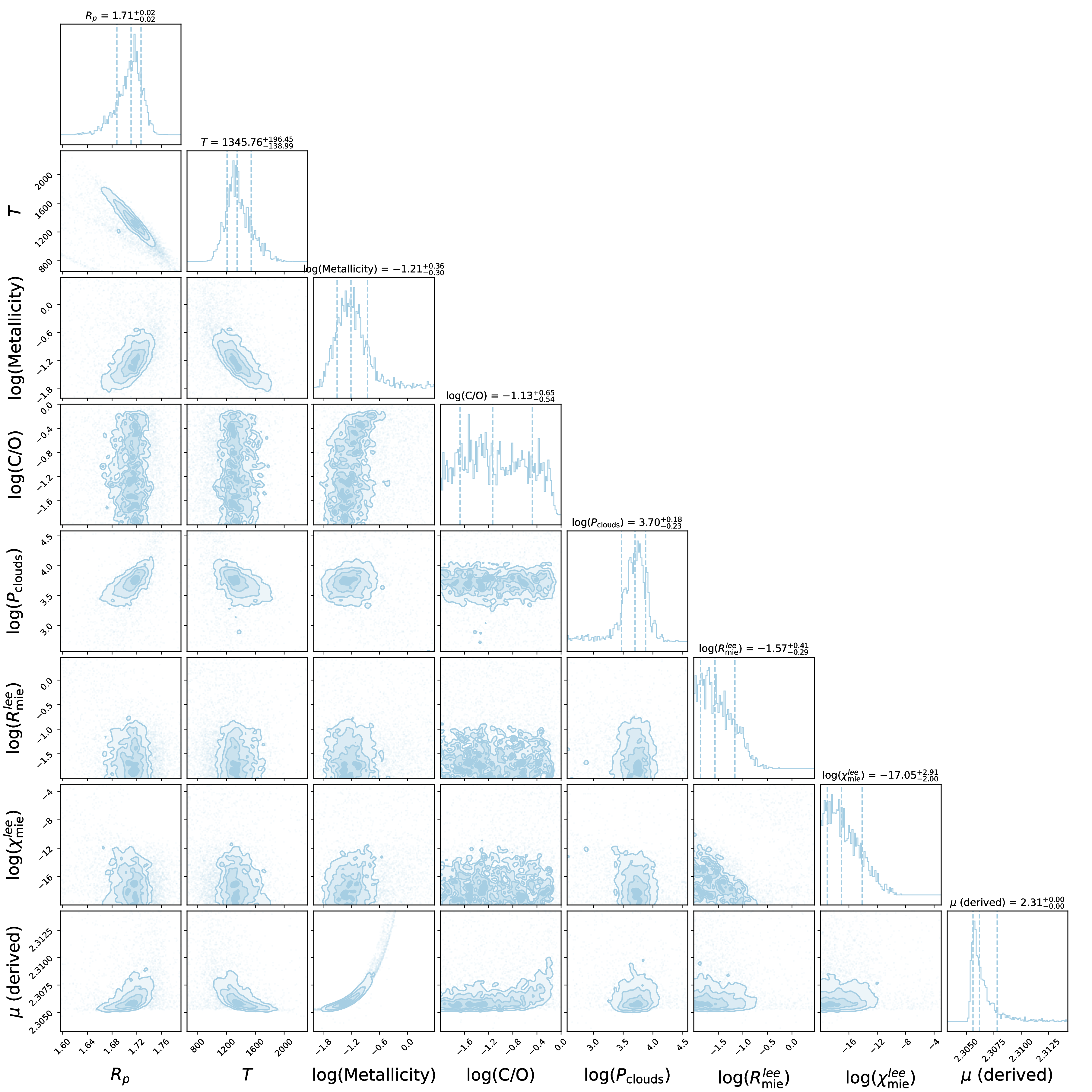}
    \caption{Posterior distribution figure for WASP-17b corresponding to the best-fit model with ACE equilibrium chemistry and clouds and hazes.}
    \label{fig:post_w17}
\end{figure*}

\newpage
\subsection{WASP-39b}
\label{section:appendix_w39}

\begin{figure*}[h!]
    \centering
    \includegraphics[width=\textwidth]{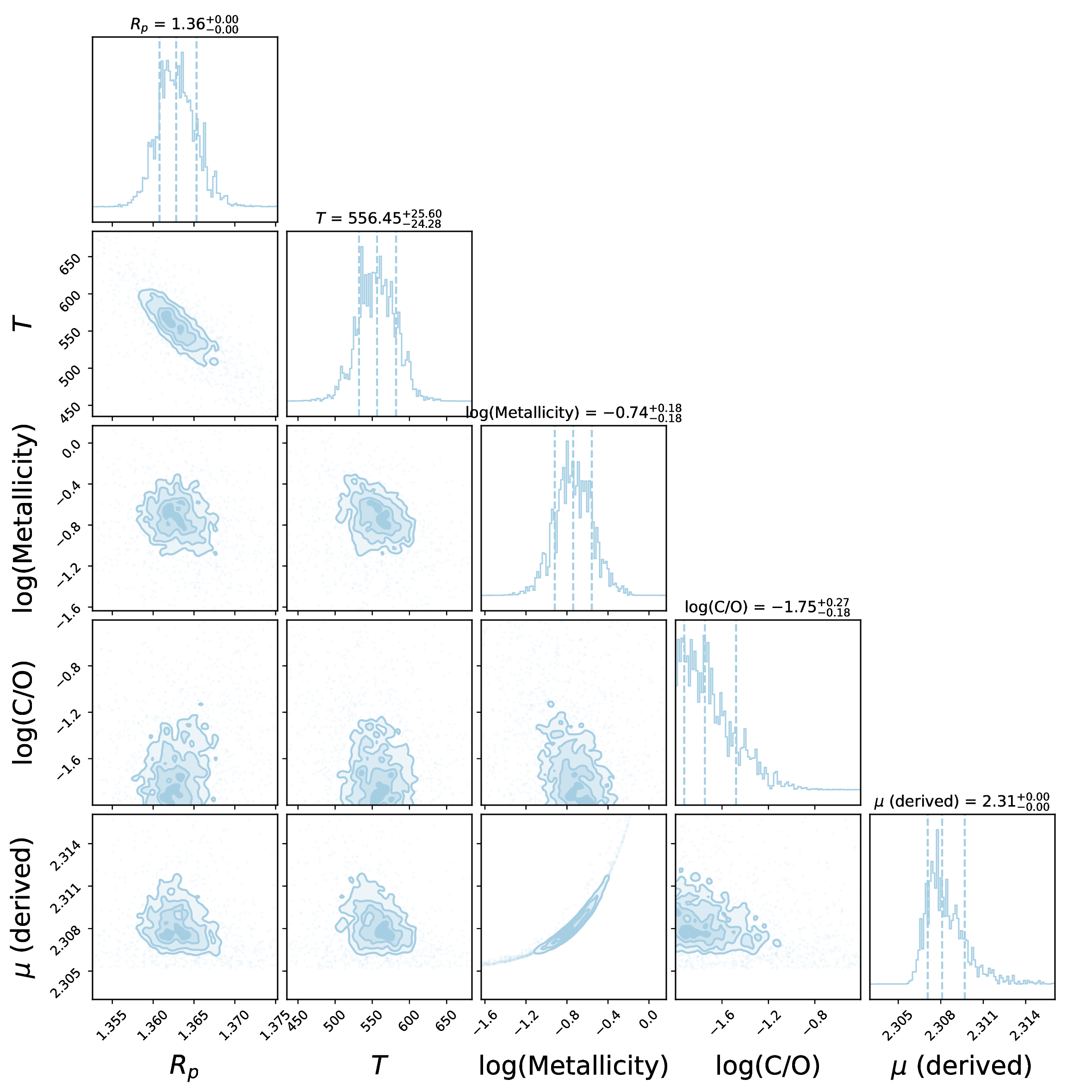}
    \caption{Posterior distribution figure for WASP-39b corresponding to the best-fit model with ACE equilibrium chemistry and neither clouds nor hazes.}
    \label{fig:post_w39}
\end{figure*}

\end{appendix}

\end{document}